\documentclass[12pt,a4paper]{article}
\makeatletter
\def\eqnarray{%
\stepcounter{equation}%
\let\@currentlabel=\theequation
\global\@eqnswtrue
\global\@eqcnt\z@
\tabskip\@centering
\let\\=\@eqncr
$$\halign to \displaywidth\bgroup\@eqnsel\hskip\@centering
$\displaystyle\tabskip\z@{##}$&\global\@eqcnt\@ne
\hfil$\displaystyle{{}##{}}$\hfil
&\global\@eqcnt\tw@$\displaystyle\tabskip\z@{##}$\hfil
\tabskip\@centering&\llap{##}\tabskip\z@\cr}
\makeatother
\newcommand{\fukuso}{{\mathbf C}}

\begin{document}

\title{\sl Explicit Form of Evolution Operator of Three Atoms Tavis--Cummings 
Model}
\author{
  Kazuyuki FUJII
  \thanks{E-mail address : fujii@yokohama-cu.ac.jp },\quad 
  Kyoko HIGASHIDA 
  \thanks{E-mail address : s035577d@yokohama-cu.ac.jp },\quad 
  Ryosuke KATO 
  \thanks{E-mail address : s035559g@yokohama-cu.ac.jp }\\
  Tatsuo SUZUKI
  \thanks{E-mail address : suzukita@gm.math.waseda.ac.jp },\quad 
  Yukako WADA 
  \thanks{E-mail address : s035588a@yokohama-cu.ac.jp }\\
  ${}^{*,\dagger,\ddagger,\P}$Department of Mathematical Sciences\\
  Yokohama City University, 
  Yokohama, 236--0027, 
  Japan\\
  ${}^\S$Department of Mathematical Sciences\\
  Waseda University, 
  Tokyo, 169--8555, 
  Japan\\
  }
\date{}
\maketitle
%
%
%
%
\begin{abstract}
  In this letter the explicit form of evolution operator of three atoms 
  Tavis--Cummings model is given, which is a generalization of the paper 
  quant-ph/0403008. 
\end{abstract}
%


%
%
%
%


The purpose of this letter is to give an explicit form to the evolution 
operator of Tavis--Cummings model (\cite{TC}) with some atoms. 
This model is a very important one in Quantum Optics and has been studied 
widely, see \cite{books} as general textbooks in quantum optics. 

We are studying a quantum computation and therefore want to study the model 
from this point of view, namely the quantum computation based on atoms of 
laser--cooled and trapped linearly in a cavity. We must in this model 
construct a controlled NOT gate or other controlled unitary gates to perform 
a quantum computation, see \cite{KF1} as a general introduction to 
this subject. 

For that aim we need the explicit form of evolution operator of the models 
with one, two and three atoms (at least). As to the model of one atom or 
two atoms it is more or less known (see \cite{papers}), while as to the case 
of three atoms it has not been given as far as we know. 
Since we succeeded in finding the explicit form for three atoms case 
we report it \footnote{J.C.Retamal et al might have obtained the same result 
by another method \cite{JCR}}. 

The Tavis--Cummings model (with $n$--atoms) that we will treat in this paper 
can be written as follows (we set $\hbar=1$ for simplicity). 
\begin{equation}
\label{eq:hamiltonian}
H=
\omega {1}_{L}\otimes a^{\dagger}a + 
\frac{\Delta}{2} \sum_{i=1}^{n}\sigma^{(3)}_{i}\otimes {\bf 1} +
g\sum_{i=1}^{n}\left(
\sigma^{(+)}_{i}\otimes a+\sigma^{(-)}_{i}\otimes a^{\dagger} \right),
\end{equation}
where $\omega$ is the frequency of radiation field, $\Delta$ the energy 
difference of two level atoms, $a$ and $a^{\dagger}$ are 
annihilation and creation operators of the field, and $g$ a coupling constant, 
and $L=2^{n}$. Here $\sigma^{(+)}_{i}$, $\sigma^{(-)}_{i}$ and 
$\sigma^{(3)}_{i}$ are given as 
\begin{equation}
\sigma^{(s)}_{i}=
1_{2}\otimes \cdots \otimes 1_{2}\otimes \sigma_{s}\otimes 1_{2}\otimes \cdots 
\otimes 1_{2}\ (i-\mbox{position})\ \in \ M(L,\fukuso)
\end{equation}
where $s$ is $+$, $-$ and $3$ respectively and 
\begin{equation}
\label{eq:sigmas}
\sigma_{+}=
\left(
  \begin{array}{cc}
    0& 1 \\
    0& 0
  \end{array}
\right), \quad 
\sigma_{-}=
\left(
  \begin{array}{cc}
    0& 0 \\
    1& 0
  \end{array}
\right), \quad 
\sigma_{3}=
\left(
  \begin{array}{cc}
    1& 0  \\
    0& -1
  \end{array}
\right), \quad 
1_{2}=
\left(
  \begin{array}{cc}
    1& 0  \\
    0& 1
  \end{array}
\right).
\end{equation}

Here let us rewrite the hamiltonian (\ref{eq:hamiltonian}). If we set 
\begin{equation}
\label{eq:large-s}
S_{+}=\sum_{i=1}^{n}\sigma^{(+)}_{i},\quad 
S_{-}=\sum_{i=1}^{n}\sigma^{(-)}_{i},\quad 
S_{3}=\frac{1}{2}\sum_{i=1}^{n}\sigma^{(3)}_{i},
\end{equation}
then (\ref{eq:hamiltonian}) can be written as 
\begin{equation}
\label{eq:hamiltonian-2}
H=
\omega {1}_{L}\otimes a^{\dagger}a + \Delta S_{3}\otimes {\bf 1} + 
g\left(S_{+}\otimes a + S_{-}\otimes a^{\dagger} \right)
\equiv H_{0}+V,
\end{equation}
which is very clear. We note that $\{S_{+},S_{-},S_{3}\}$ satisfy the 
$su(2)$--relation 
\begin{equation}
[S_{3},S_{+}]=S_{+},\quad [S_{3},S_{-}]=-S_{-},\quad [S_{+},S_{-}]=2S_{3}.
\end{equation}
However, the representation $\rho$ defined by 
$
\rho(\sigma_{+})=S_{+},\ \rho(\sigma_{-})=S_{-},\ 
\rho(\sigma_{3}/2)=S_{3}
$
is a reducible representation of $su(2)$. 

We would like to solve the Schr{\" o}dinger equation 
\begin{equation}
\label{eq:schrodinger}
i\frac{d}{dt}U=HU=\left(H_{0}+V\right)U, 
\end{equation}
where $U$ is a unitary operator (called the evolution operator). 
We can solve this equation by using the {\bf method of constant variation}. 
The result is well--known to be 
\begin{equation}
\label{eq:full-solution}
U(t)=\left(\mbox{e}^{-it\omega S_{3}}\otimes 
\mbox{e}^{-it\omega a^{\dagger}a}\right)
\mbox{e}^{-itg\left(S_{+}\otimes a + S_{-}\otimes a^{\dagger}\right)}
\end{equation}
under the resonance condition $\Delta=\omega$, 
where we have dropped the constant unitary operator for simplicity. 
Therefore 
we have only to calculate the term (\ref{eq:full-solution}) explicitly, 
which is however a very hard task \footnote{the situation is very similar to 
that of the paper quant-ph/0312060 in \cite{qudit-papers}}. 
In the following we set 
\begin{equation}
\label{eq:A}
A=S_{+}\otimes a + S_{-}\otimes a^{\dagger}
\end{equation}
for simplicity. 
We can determine\ $\mbox{e}^{-itgA}$\ for $n=1$ (one atom case), 
$n=2$ (two atoms case) and $n=3$ (three atoms case) completely. 

\vspace{3mm}
\par \noindent 
{\bf One Atom Case}\quad In this case $A$ in (\ref{eq:A}) is written as 
\begin{equation}
\label{eq:A-one}
A_{1}=
\left(
  \begin{array}{cc}
    0&           a \\
    a^{\dagger}& 0
  \end{array}
\right).
\end{equation}
By making use of the relation 
\begin{equation}
\label{eq:relation-one}
{A_{1}}^{2}=
\left(
  \begin{array}{cc}
    aa^{\dagger}&   0          \\
    0           & a^{\dagger}a 
  \end{array}
\right)=
\left(
  \begin{array}{cc}
    N+1& 0  \\
    0  & N
  \end{array}
\right)
\end{equation}
with the number operator $N$ 
we have 
\begin{equation}
\label{eq:solution-one}
\mbox{e}^{-itgA_{1}}=
\left(
  \begin{array}{cc}
  \mbox{cos}\left(tg\sqrt{N+1}\right)& 
  -i\frac{\mbox{sin}\left(tg\sqrt{N+1}\right)}{\sqrt{N+1}}a  \\
  -i\frac{\mbox{sin}\left(tg\sqrt{N}\right)}{\sqrt{N}}a^{\dagger}& 
  \mbox{cos}\left(tg\sqrt{N}\right)
  \end{array}
\right).
\end{equation}
We obtained the explicit form of solution. However, this form is more or less 
well--known, see for example the second book in \cite{books}. 
We note that (\ref{eq:solution-one}) can be decomposed as 
\begin{eqnarray}
&&\left(
  \begin{array}{cc}
  \mbox{cos}\left(tg\sqrt{N+1}\right)& 
  -i\frac{\mbox{sin}\left(tg\sqrt{N+1}\right)}{\sqrt{N+1}}a  \\
  -i\frac{\mbox{sin}\left(tg\sqrt{N}\right)}{\sqrt{N}}a^{\dagger}& 
  \mbox{cos}\left(tg\sqrt{N}\right)
  \end{array}
\right)       \nonumber \\
=&&
\left(
  \begin{array}{cc}
  1 & 0 \\
  -i\frac{\mbox{tan}\left(tg\sqrt{N}\right)}{\sqrt{N}}a^{\dagger} & 1
  \end{array}
\right)
\left(
  \begin{array}{cc}
  \mbox{cos}\left(tg\sqrt{N+1}\right) & 0 \\
  0 & \frac{1}{\mbox{cos}\left(tg\sqrt{N}\right)}
  \end{array}
\right)
\left(
  \begin{array}{cc}
  1 & -i\frac{\mbox{tan}\left(tg\sqrt{N+1}\right)}{\sqrt{N+1}}a \\
  0 & 1
  \end{array}
\right).        \nonumber \\
&&
\end{eqnarray}

This is a Gauss decomposition of unitary operator. This may be used to 
construct a theory of ``quantum" representation of a non--commutative group, 
which is now under consideration.

\vspace{3mm}
\par \noindent 
{\bf Two Atoms Case}\quad In this case $A$ in (\ref{eq:A}) is written as 
\begin{equation}
\label{eq:A-two}
A_{2}=
\left(
  \begin{array}{cccc}
    0 &          a & a &           0  \\
    a^{\dagger}& 0 & 0 &           a  \\
    a^{\dagger}& 0 & 0 &           a  \\
    0 & a^{\dagger}& a^{\dagger} & 0
  \end{array}
\right).
\end{equation}

Our method is to reduce the $4\times 4$--matrix $A_{2}$ in (\ref{eq:A-two}) to 
a $3\times 3$--matrix $B_{1}$ in the following to make our calculation 
easier. 
For that aim we prepare the following matrix
\[
T=
\left(
  \begin{array}{cccc}
    0 &   1                & 0                  & 0   \\
    \frac{1}{\sqrt{2}} & 0 & \frac{1}{\sqrt{2}} & 0   \\
   -\frac{1}{\sqrt{2}} & 0 & \frac{1}{\sqrt{2}} & 0   \\
    0 &   0                &   0                  & 1
  \end{array}
\right),
\]
then it is easy to see 
\[
T^{\dagger}A_{2}T=
\left(
  \begin{array}{cccc}
    0  &                     &                     &            \\
       & 0                   & \sqrt{2}a           & 0          \\
       & \sqrt{2}a^{\dagger} & 0                   & \sqrt{2}a  \\
       & 0                   & \sqrt{2}a^{\dagger} & 0
  \end{array}
\right)\equiv 
\left(
  \begin{array}{cc}
     0 &       \\
       & B_{1} 
  \end{array}
\right)
\]
where 
$
B_{1}=J_{+}\otimes a + J_{-}\otimes a^{\dagger}
$
and $\left\{J_{+},J_{-}\right\}$ are just generators of (spin one) 
irreducible representation of (\ref{eq:sigmas}). We note that this means 
a well--known decomposition of spin 
$\frac{1}{2}\otimes \frac{1}{2}=0\oplus 1$. 

Therefore to calculate $\mbox{e}^{-itgA_{2}}$ we have only to do 
$\mbox{e}^{-itgB_{1}}$. 
Noting the relation 
\[
{B_{1}}^{3}=
\left(
  \begin{array}{ccc}
    2(2N+3) &         &          \\
            & 2(2N+1) &          \\
            &         & 2(2N-1)
  \end{array}
\right)B\equiv DB_{1},
\]
we obtain 
\begin{equation}
\label{eq:solution-two-more(reduced)}
\mbox{e}^{-itgB_{1}}=
\left(
  \begin{array}{ccc}
    b_{11} & b_{12} & b_{13} \\
    b_{21} & b_{22} & b_{23} \\
    b_{31} & b_{32} & b_{33}
  \end{array}
\right)
\end{equation}
where 
\begin{eqnarray}
b_{11}&=&\frac{N+2+(N+1)\mbox{cos}\left(tg\sqrt{2(2N+3)}\right)}{2N+3},\quad 
b_{12}=-i\frac{\mbox{sin}\left(tg\sqrt{2(2N+3)}\right)}{\sqrt{2N+3}}a,
\nonumber \\
b_{13}&=&\frac{-1+\mbox{cos}\left(tg\sqrt{2(2N+3)}\right)}{2N+3}a^{2},\quad 
b_{21}=
-i\frac{\mbox{sin}\left(tg\sqrt{2(2N+1)}\right)}{\sqrt{2N+1}}a^{\dagger},
\nonumber \\
b_{22}&=&\mbox{cos}\left(tg\sqrt{2(2N+1)}\right),\quad 
b_{23}=-i\frac{\mbox{sin}\left(tg\sqrt{2(2N+1)}\right)}{\sqrt{2N+1}}a, 
\nonumber \\
b_{31}&=&
\frac{-1+\mbox{cos}\left(tg\sqrt{2(2N-1)}\right)}{2N-1}{(a^{\dagger})^2},
\quad 
b_{32}=-i\frac{\mbox{sin}\left(tg\sqrt{2(2N-1)}\right)}{\sqrt{2N-1}}
a^{\dagger},
\nonumber \\
b_{33}&=&\frac{N-1+N\mbox{cos}\left(tg\sqrt{2(2N-1)}\right)}{2N-1}. \nonumber
\end{eqnarray}

\vspace{3mm}
\par \noindent 
{\bf Three Atoms Case}\quad In this case $A$ in (\ref{eq:A}) is written as 
\begin{equation}
\label{eq:A-three}
A_{3}=
\left(
  \begin{array}{cccccccc}
    0 &          a & a &           0  & a & 0 & 0 & 0          \\
    a^{\dagger}& 0 & 0 &           a  & 0 & a & 0 & 0          \\
    a^{\dagger}& 0 & 0 &           a  & 0 & 0 & a & 0          \\
    0 & a^{\dagger}& a^{\dagger} & 0  & 0 & 0 & 0 & a          \\
    a^{\dagger}& 0 & 0  &  0          & 0 & a & a & 0          \\
    0 & a^{\dagger}& 0  & 0   & a^{\dagger} &  0 & 0 & a       \\
    0 & 0 & a^{\dagger} & 0  & a^{\dagger} &  0 & 0 & a        \\
    0 & 0 & 0 & a^{\dagger} & 0 & a^{\dagger} & a^{\dagger} & 0    
  \end{array}
\right).
\end{equation}

We would like to look for the explicit form of solution like 
(\ref{eq:solution-one}) or (\ref{eq:solution-two-more(reduced)}). 
If we set 
\[
T=
\left(
  \begin{array}{cccccccc}
    0 & 0 & 0 & 0 & 1 & 0 & 0 & 0 \\
    \frac{1}{\sqrt{2}} & 0 & \frac{1}{\sqrt{6}} & 0 & 0 & 
    \frac{1}{\sqrt{3}} & 0 & 0 \\
    \frac{-1}{\sqrt{2}} & 0 & \frac{1}{\sqrt{6}} & 0 & 0 & 
    \frac{1}{\sqrt{3}} & 0 & 0 \\ 
    0 & 0 & 0 & \frac{2}{\sqrt{3}} & 0 & 0 & \frac{1}{\sqrt{3}} & 0  \\
    0 & 0 & \frac{-2}{\sqrt{3}} & 0 & 0 & \frac{1}{\sqrt{3}} & 0 & 0 \\
    0 & \frac{1}{\sqrt{2}} & 0 & \frac{-1}{\sqrt{6}} & 0 & 0 & 
    \frac{1}{\sqrt{3}} & 0 \\
    0 & \frac{-1}{\sqrt{2}} & 0 & \frac{-1}{\sqrt{6}} & 0 & 0 & 
    \frac{1}{\sqrt{3}} & 0 \\
    0 & 0 & 0 & 0 & 0 & 0 & 0 & 1
  \end{array}
\right),
\]
then it is not difficult to see 
\[
T^{\dagger}A_{3}T=
\left(
  \begin{array}{cccccccc}
     0 & a &   &   &   &   &   &                       \\
    a^{\dagger}& 0 &   &   &    &   &   &              \\
       &   & 0 &  a &   &   &    &                     \\
       &   & a^{\dagger} & 0 &   &   &   &             \\
       &   &   &   & 0 & \sqrt{3}a & 0 & 0             \\
       &   &   &   & \sqrt{3}a^{\dagger} & 0 & 2a & 0  \\
       &   &   &   & 0 & 2a^{\dagger} & 0 & \sqrt{3}a  \\   
       &   &   &   & 0 & 0 & \sqrt{3}a^{\dagger} & 0
  \end{array}
\right)\equiv 
\left(
  \begin{array}{ccc}
     A_{1} &       &        \\
           & A_{1} &        \\ 
           &       & B_{2}
  \end{array}
\right).
\]
This means a decomposition of spin $\frac{1}{2}\otimes \frac{1}{2}\otimes 
\frac{1}{2}=\frac{1}{2}\oplus \frac{1}{2}\oplus \frac{3}{2}$. 
Therefore we have only to calculate $\mbox{e}^{-itgB_{2}}$, which is however 
not easy. The result is 
\begin{eqnarray}
&&\mbox{e}^{-itgB_{2}}    \nonumber \\
=&&
\left(
  \begin{array}{cccc}
    f_{1}(N+2) & -\sqrt{3}ih_{2}(N+2)a & 2\sqrt{3}f_{3}(N+2)a^{2} & 
    -6ih_{3}(N+2)a^{3}   \\
    -\sqrt{3}ih_{2}(N+1)a^{\dagger} & f_{2}(N+1) & -2i\tilde{h}_{3}(N+1)a & 
    2\sqrt{3}f_{3}(N+1)a^{2} \\
    2\sqrt{3}f_{3}(N)(a^{\dagger})^{2} & -2i\tilde{h}_{3}(N)a^{\dagger} & 
    f_{4}(N) & -\sqrt{3}ih_{4}(N)a  \\
    -6ih_{3}(N-1)(a^{\dagger})^{3} & 2\sqrt{3}f_{3}(N-1)(a^{\dagger})^{2} & 
    -\sqrt{3}ih_{4}(N-1)a^{\dagger} & f_{5}(N-1)   
  \end{array}
\right)                  \nonumber \\ 
&{}& 
\end{eqnarray}
where 
\begin{eqnarray}
f_{1}(N)&=&\left\{v_{+}(N)\mbox{cos}(tg\sqrt{\lambda_{+}(N)})-
v_{-}(N)\mbox{cos}(tg\sqrt{\lambda_{-}(N)})\right\}/(2\sqrt{d(N)}), 
\nonumber \\
f_{2}(N)&=&\left\{w_{+}(N)\mbox{cos}(tg\sqrt{\lambda_{+}(N)})-
w_{-}(N)\mbox{cos}(tg\sqrt{\lambda_{-}(N)})\right\}/(2\sqrt{d(N)}), 
\nonumber \\
h_{2}(N)&=&\left\{\frac{w_{+}(N)}{\sqrt{\lambda_{+}(N)}}
\mbox{sin}(tg\sqrt{\lambda_{+}(N)})-
\frac{w_{-}(N)}{\sqrt{\lambda_{-}(N)}}
\mbox{sin}(tg\sqrt{\lambda_{-}(N)})\right\}/(2\sqrt{d(N)}), 
\nonumber \\
f_{3}(N)&=&\left\{\mbox{cos}(tg\sqrt{\lambda_{+}(N)})-
\mbox{cos}(tg\sqrt{\lambda_{-}(N)})\right\}/(2\sqrt{d(N)}), 
\nonumber \\
h_{3}(N)&=&\left\{\frac{1}{\sqrt{\lambda_{+}(N)}}
\mbox{sin}(tg\sqrt{\lambda_{+}(N)})-
\frac{1}{\sqrt{\lambda_{-}(N)}}
\mbox{sin}(tg\sqrt{\lambda_{-}(N)})\right\}/(2\sqrt{d(N)}), 
\nonumber \\
f_{4}(N)&=&\left\{v_{+}(N)\mbox{cos}(tg\sqrt{\lambda_{-}(N)})-
v_{-}(N)\mbox{cos}(tg\sqrt{\lambda_{+}(N)})\right\}/(2\sqrt{d(N)}), 
\nonumber \\
h_{4}(N)&=&\left\{\frac{v_{+}(N)}{\sqrt{\lambda_{-}(N)}}
\mbox{sin}(tg\sqrt{\lambda_{-}(N)})-
\frac{v_{-}(N)}{\sqrt{\lambda_{+}(N)}}
\mbox{sin}(tg\sqrt{\lambda_{+}(N)})\right\}/(2\sqrt{d(N)}), 
\nonumber \\
f_{5}(N)&=&\left\{w_{+}(N)\mbox{cos}(tg\sqrt{\lambda_{-}(N)})-
w_{-}(N)\mbox{cos}(tg\sqrt{\lambda_{+}(N)})\right\}/(2\sqrt{d(N)}), 
\nonumber \\
\tilde{h}_{3}(N)&=&\left\{\sqrt{\lambda_{+}(N)}
\mbox{sin}(tg\sqrt{\lambda_{+}(N)})-
\sqrt{\lambda_{-}(N)}
\mbox{sin}(tg\sqrt{\lambda_{-}(N)})\right\}/(2\sqrt{d(N)})
\nonumber 
\end{eqnarray}
and 
\[
d(N)=16N^{2}+9,
\lambda_{\pm}=5N\pm \sqrt{d(N)}, 
v_{\pm}=-2N-3\pm \sqrt{d(N)}, 
w_{\pm}=2N-3\pm \sqrt{d(N)}. 
\]

\vspace{5mm}
We obtained the explicit form of evolution operator of the Tavis--Cummings 
model for three atoms case, so there are many applications to quantum optics 
or mathematical physics, see for example \cite{papers}. 
In the near future we will apply the result to a quantum computation based on 
atoms of laser--cooled and trapped linearly in a cavity \cite{FHKW}. 

We conclude this paper by making a comment. The Tavis--Cummings model 
is based on (only) two energy levels of atoms. However, an atom has in general 
infinitely many energy levels, so it is natural to use this possibility. 
We are also studying a quantum computation based on multi--level systems of 
atoms (a qudit theory) \cite{qudit-papers}. Therefore we would like to extend 
the Tavis--Cummings model based on two--levels to a model based on 
multi--levels. This is a very challenging task.


\end{document}